\documentclass[a4paper,11pt,onecolumn,superscriptaddress]{article}
\usepackage{comment}
\usepackage{graphicx}
\usepackage{epsf}
\usepackage{bm}
\usepackage{amsmath}
\usepackage{amsfonts}
\usepackage{amssymb}
\usepackage{epstopdf}
\usepackage{color}
\usepackage[dvipsnames]{xcolor}
\usepackage{verbatim}
\usepackage{multirow}
\usepackage{soul}
\usepackage{bm}

\usepackage[width=0.00cm, height=0.00cm, left=1.50cm, right=1.50cm, top=2.00cm, bottom=2.00cm]{geometry}
\usepackage{microtype}
\usepackage[colorlinks = true,
            linkcolor = teal,
            urlcolor  = teal,
            citecolor = teal,
            anchorcolor = blue]{hyperref}
\usepackage[capitalize]{cleveref}
\usepackage[normalem]{ulem}
\usepackage{enumitem}
\usepackage{booktabs}

\usepackage{lipsum}

\makeatletter\let\expandableinput\@@input\makeatother

\begin{document}

\begin{center}
		\vspace{0.4cm} {\large{\bf Testing $f(T)$ Gravity with Cosmological Observations: Confronting the Hubble Tension and Implications for the Late-Time Universe }} \\
		\vspace{0.4cm}
		\normalsize{ Saurabh Verma $^1$, Archana Dixit$^2$, Anirudh Pradhan$^3$, M. S. Barak$^4$ }\\
		\vspace{5mm}
		\normalsize{$^{1,4}$ Department of Mathematics, Indira Gandhi University, Meerpur, Haryana 122502, India\\
		\normalsize{$^{2 }$ Department of Mathematics, Gurugram University, Gurugram, Harayana, India}\\ 
	
		\normalsize{$^{3 }$ Centre for Cosmology, Astrophysics and Space Science (CCASS), GLA University, Mathura, Uttar Pradesh, India}\\ 

		\vspace{2mm}
		$^1$Email address: saurabh.math.rs@igu.ac.in\\
		$^2$Email address: archana.ibs.maths@gmail.com\\
		$^3$Email address: anirudh.pradhan@gmail.com\\
        $^4$Email address: ms$_{-}$barak@igu.ac.in\\}
\end{center}

{\noindent {\bf Abstract.}

In recent years, modifications to General Relativity (GR) have been explored to address cosmological observations, particularly in the context of late-time cosmic acceleration. Among these, modifications based on the Teleparallel Equivalent of General Relativity (TEGR), particularly $f(T)$ gravity, have gained significant attention. In this work, we investigate the scalar perturbations in $f(T)$ gravity, focusing on how these perturbations modify the Poisson and lensing equations and how they impact cosmological observables. By incorporating observational data from cosmic chromatometers, Big Bang nucleosynthesis, the DESI BAO survey, and Type Ia Supernovae (SNe Ia), we derive constraints on the parameters of the $f(T)$ power-law model. Our results suggest that $f(T)$ gravity can effectively alleviate some of the tensions observed in the standard $\Lambda$CDM model, including the Hubble constant ($H_0$) discrepancy. Furthermore, the evolution of the supernova luminosity and its dependence on the gravitational constant are considered to refine the measurement of cosmological parameters. The model's ability to address the $H_0$ tension is critically examined, and we find that $f(T)$ gravity offers a viable alternative to the standard model. The work concludes by comparing the fits of the $f(T)$ gravity model to the $\Lambda$CDM model using various information criteria, revealing key insights into the viability of modified gravity in contemporary cosmology.\\
\smallskip 
{\bf Keywords}: f(T) gravity, Cosmological observations, Hubble tension, Accelerating Universe\\
PACS: 98.80.-k 
\section{Introduction}

General Relativity (GR) has undergone various extensions to better account for cosmological and astrophysical observations (see reviews in \cite{ref1,ref2,ref3,ref4,ref5}). These alternative theories of gravity have been extensively analyzed, particularly those  introduce additional gravitational degrees of freedom. Such modifications are capable of driving the Universe’s accelerated expansion in its later stages and offer frameworks to extend the standard $\Lambda$CDM model. However, many of these extensions yield similar observational consequences, resulting in a theoretical degeneracy among the proposed models. Recently, there has been growing interest in torsion-based approaches, particularly those stemming from the Teleparallel Equivalent of General Relativity (TEGR) \cite{ref6}, as potential foundations for modified gravity (MG) theories. In this context, the simplest extension—known as $f(T)$ gravity—arises by generalizing the TEGR Lagrangian from the torsion scalar $T$ to an arbitrary function of $T$ (for overviews, see \cite{ref7,ref8,ref9}).In recent years, modifications of general relativity (GR) have gained significant attention as possible explanations for the accelerating expansion of the universe. Among these, teleparallel gravity and specifically its extended formulation, known as \textcolor{red}{$f(T)$} gravity, has emerged as a promising alternative. Unlike standard GR, which is based on the curvature of spacetime as described by the Levi-Civita connection, teleparallel gravity replaces curvature with torsion, using the Weitzenböck connection. This alternative formulation provides a different theoretical perspective on gravity while maintaining consistency with observational data.The fundamental idea behind gravity is to generalize the teleparallel equivalent of general relativity (TEGR).
This approach is analogous to gravity, where the Ricci scalar is replaced with a function . The modifications in gravity lead to different field equations and dynamics, which can potentially address cosmological issues such as dark energy and late-time cosmic acceleration without invoking an explicit cosmological constant.\\

Modified theories of gravity have attracted significant attention in recent years, driven by two primary motivations beyond the well-established success of the standard cosmological model based on General Relativity (GR). The first motivation lies in seeking alternative explanations for the late-time acceleration of the Universe without invoking unknown dark components. The second stems from theoretical efforts to overcome the limitations of GR by achieving a renormalizable and quantum-consistent theory of gravity. Among these approaches, teleparallel gravity has emerged as a promising classical framework for cosmological modeling, particularly due to its reformulation of gravity using torsion instead of curvature. In its extended form, 
$f(T)$ gravity maintains second-order field equations while allowing richer cosmological dynamics. While not a quantum theory of gravity, its gauge structure and simplicity make it an attractive alternative to GR in describing the late-time acceleration of the universe. \\

Nojiri and Odintsov \cite{ref9a}) proposed modified gravity as an alternative explanation to dark energy in cosmological theories. De Andrade \cite{ref9b})provided foundational insights into the teleparallel formulation equivalent to general relativity. Dagwal and Pawar \cite{ref9c}) analyzed cosmological models involving two fluids within the scope of $f(T)$ gravity. Basilakos \cite{ref9d}) conducted a detailed investigation into Noether symmetries and the resulting analytical solutions in $f(T)$ cosmology. Additionally, Cai et al. \cite{ref9})contributed notably by exploring diverse cosmological scenarios under $f(T)$ teleparallel gravity. Pawar et al.\cite{ref9e}) further studied the flat Friedmann–Robertson–Walker (FRW) model, incorporating dual energy components within the context of fractal cosmology.\\
 
The $\Lambda$CDM model has emerged as the most widely accepted framework for explaining the large-scale evolution of the Universe, particularly following the discovery of its late-time accelerated expansion \cite{ref9f,ref9g}. This model incorporates a cosmological constant ($\Lambda$), representing dark energy -a mysterious component with negative pressure — directly into Einstein’s field equations. Additionally, it assumes the dominance of cold dark matter (CDM), a form of pressureless, non-baryonic matter, which constitutes the majority of the matter content in galaxies. The $\Lambda$CDM model provides an excellent fit to a wide range of observational data, from type Ia supernovae to the cosmic microwave background (CMB) temperature anisotropies \cite{ref9h}. Nonetheless, recent observations of the Universe’s late-time behavior have raised questions about the completeness of general relativity (GR) in describing gravitational phenomena, sparking interest in exploring extensions or modifications to the standard cosmological model, such as modified gravity theories \cite{ref9i,ref9j}. Among these alternatives, several viable forms of the $f(T)$ function have been proposed in the literature \cite{ref9k,ref9l}. In particular, studies in Refs. \cite{ref9N,ref9N1,ref9N2,ref9N3} have performed statistical analyzes to constrain the cosmological parameters of these models using various observational data sets.However, agreement on the $\Lambda$CDM model is currently being challenged by increasingly precise measurements of the cosmological parameters \cite{ref10}. As measured by the Hubble constant, or $H_0$, the observed value of the current cosmic expansion rate is undoubtedly the most substantial conflict with the standard model prediction. As is well known, Planck-CMB data analysis yields $H_0 =67.4\pm0.5$ $\rm km \rm s^{-1} \rm Mpc^{-1}$ \cite{ref11}, given the minimal $\Lambda$CDM scenario. 
This result shows a discrepancy at approximately the $5\sigma$ level when compared to the local measurement of $H_0 = 73.30 \pm 1.04$ $\rm km\ s^{-1}\ Mpc^{-1}$ reported by the SH0ES team \cite{ref12}.\\

A larger value of the Hubble constant aligns with several other observations made at later cosmic times \cite{ref10,ref13,ref14}. These observational discrepancies have sparked extensive discussions in the literature, as they are unlikely to be solely attributed to numerous independent systematic errors. This has led to speculation that resolving the so-called $H_0$ tension may require physics beyond the standard cosmological model (for comprehensive reviews,  \cite{ref10,ref13,ref14,ref15}). Since the SH0ES determination of $H_0$ directly depends on estimates of the supernova absolute magnitude $M_B$, some have argued that the real issue might lie in a discrepancy in $M_B$ itself \cite{ref16,ref17}. Specifically, SH0ES reports a value of $M_B = -19.244 \pm 0.037$ mag \cite{ref17}, while the inverse distance ladder approach, using cosmic microwave background (CMB) constraints, yields $M_B = -19.401 \pm 0.027$ mag \cite{ref18}. This results in a tension of approximately $3.4\sigma$. As discussed in \cite{ref16,ref17,ref19,ref20}, Cepheid-based calibrations are primarily used to assess this tension in supernova luminosity rather than to directly resolve the $H_0$ discrepancy. Additionally, attempts to address the $H_0$ tension through modifications of General Relativity (GR) have also been explored in recent studies \cite{ref10,ref13,ref14}.\\

Allowing the gravitational constant to vary with redshift $z$ introduces significant changes in the context of modified gravity (MG) theories, potentially affecting the astrophysical characteristics of Type Ia supernovae (SNe Ia)\cite{ref21,ref22}. This phenomenon has been both investigated and utilized to place limits on different theoretical models by considering how the intrinsic luminosity evolves due to a changing Newtonian constant $G$ \cite{ref23,ref24,ref25,ref26}. Because of this sensitivity, it is crucial to consider the role of MG effects on SNe Ia when inferring cosmological parameters. One of the primary aims of this study is to incorporate these luminosity evolution adjustments into the framework of 
$f(T)$ gravity and to examine their implications. It is well known that general relativity (GR) has a substantial influence on key cosmological observables, including the development of density perturbations, the formation of large-scale structures, anisotropies in the cosmic microwave background (CMB), and other signals across the Universe\cite{ref1,ref2}). These deviations are often described through phenomenological parameters  $\Sigma$  and $\mu$ , which represent modifications to the perturbed Einstein equations, linking gravitational lensing effects to the Newtonian potential and fluctuations in matter density \cite{ref27,ref28,ref29,ref30}. In this study, after formulating expressions for $\Sigma$   and  $\mu$  within the $ f(T)$ gravity framework, our secondary objective is to place additional observational constraints on $f(T)$ gravity by leveraging the most recent data from DESI BAO, BBN, and the Pantheon sample. This includes calculating the modified Hubble expansion rate $H(z)$ \cite{ref31,ref32,ref33,ref34,ref35,ref36,ref37,ref38,ref39,ref40,ref41,ref42,ref43,ref44,ref45,ref46,ref47}, along with incorporating current measurements of structure growth on sub-horizon scales \cite{ref4}.\\

The study of gravity is particularly relevant in cosmology, where it has been used to explore various aspects of the universe’s evolution. The theory provides alternative mechanisms for the observed acceleration of the cosmic expansion and offers insights into the nature of dark energy. Furthermore, scalar perturbations within gravity can influence structure formation, gravitational lensing, and cosmic microwave background (CMB) anisotropies, making it an essential topic for both theoretical and observational investigations.In this study, our focus is on placing new and updated constraints on different $f(T)$ gravity models by utilizing the most recently available observational data . Specifically, we make use of the latest astronomical observations, including: (1) Hubble parameter measurements derived from the differential evolution of cosmic chronometers (CC), combined with the most recent local determination of the Hubble constant $(H_{0})$, as well as data from BBN, BAO, DESI, and the Pantheon+ sample, to obtain improved limits on the $f(T)$ gravity models.This paper is structured as follows: In section II, we briefly review the $f(T)$ gravity and its cosmology, where the scalar perturbative evolutions in terms of
$\sum$ and $\mu$ are described along with one of the most-used $f(T)$ gravity power-law model. In section III, we describe
the data sets used for our analysis and the methodology adopted. In section IV, we present our main results and discussion, focusing on the power-law model. Section V discusses model comparisons.
Finally, Section VI summarizes our findings and suggests future possibilities for the research of teleparallel modified gravity models.

\section{$f(T)$ GRAVITY AND COSMOLOGY}
\subsection{$f(T)$ GRAVITY} 
In $f(T)$ gravity, as in all torsion-based theories, the vierbein fields 
$e^{\mu}_{A} $ provide an orthonormal basis for the tangent space at every point $x^{\mu}$  on the manifold.The metric then reads as follows: $ g_{\mu\nu} = \eta_{AB} e^{A}_{\mu} e^{B}_{\nu} $  (the coordinate and tangent spaces in this document are respectively spanned by Greek and Latin indices). \\

	In this work, following the common approach in cosmological applications of $f(T)$ gravity, we use the pure tetrad formulation, where the inertial spin connection is set to zero \textit{a priori}.
	More precisely, while we introduced the Weitzenb\"{o}ck connection as
	$\overset{\mathbf{w}}{\Gamma}{}^{\lambda}_{\mu\nu} = e^{\lambda}_{A} \, \partial_\nu e^{A}_{\mu},$
	in the fully covariant formulation this connection should include the inertial spin connection $\omega^{A}_{B\nu}$:
	\begin{equation}
		\overset{\mathbf{w}}{\Gamma}{}^{\lambda}_{\mu\nu} = e^{\lambda}_{A} \bigl( \partial_\nu e^{A}_{\mu} + \omega^{A}_{B\nu} e^{B}_{\mu} \bigr).
	\end{equation}
	Including $\omega^{A}_{B\nu}$ ensures invariance under local Lorentz transformations and keeps the theory covariant  \cite{ref48,ref49}.
	Since our study focuses on the flat FLRW background with diagonal tetrads, where the spin connection vanishes consistently, this simplification does not affect the field equations  or the derived cosmological results. 
	Based on this connection, the torsion tensor can be defined, and consequently, the torsion scalar $T$ is constructed as the Lagrangian of the teleparallel equivalent of general relativity, formed by contracting the torsion tensor as \cite{ref50,ref51}.
\begin{equation}
T \equiv \frac{1}{4} T^{\rho\mu\nu} T_{\rho\mu\nu} + \frac{1}{2} T^{\rho\mu\nu} T_{\nu\mu\rho} - T_{\rho\mu}{}^{\rho} T^{\nu\mu}{}_{\nu}. \label{2}
\end{equation}
We can extend $T$ as a function  of $f(T)$ and construct the action of $f(T)$ gravity in the following manner, inspired by the $f(R)$ extensions of general relativity.
\begin{equation}
\mathcal{S} = \frac{1}{16\pi G} \int d^{4}x e \left[ f(T) \right]. \label{3}
\end{equation}

Here we have imposed units where the speed of light is $1$, since the gravitational constant is $G$ and $e = \det(e^{A}_{\mu}) = \sqrt{-g}$. \\ 
Before deriving the field equations, we briefly discuss the theoretical basis of $f(T)$ gravity. In this approach, the teleparallel equivalent of General Relativity (TEGR), which uses the torsion scalar $T$ in place of the curvature scalar $R$, is generalized by promoting the gravitational Lagrangian to an arbitrary function $f(T)$.This extension allows the theory to go beyond GR and naturally accommodates the possibility of explaining the observed late-time cosmic acceleration without introducing an explicit cosmological constant. 
 In this work, we adopt the convention $f(T)=T$ to recover TEGR. To include a cosmological constant $\Lambda$ ($\Lambda$CDM model), this would correspond to $f(T)=T–2\Lambda$ in our notation, where $\Lambda$ is the cosmological constant and our choice is consistant.

\subsection{BACKGROUND EVOLUTION}

When applying f(T) gravity in a cosmological framework, we take into account the functional form of \textcolor{red}{$f(T)+T$}.  The matter sector (baryons and cold dark matter) must be included first, therefore the whole action is written \cite{ref52,ref53} as:
 \begin{equation}
 \mathcal{S} = \frac{1}{16\pi G} \int d^{4}x e \left[f(T)+T \right] + \mathcal{S}_{m}. \label{4}
 \end{equation}

In this work, we adopt the pure tetrad formalism of $f(T)$ gravity, assuming a vanishing inertial spin connection. This is equivalent to working in a proper frame where the antisymmetric part of the field equations vanishes. The validity of this assumption holds for the class of diagonal tetrads used in our analysis. The field equations are given by the variation of the action (4) with regard to the vierbeins as

\begin{eqnarray}
\label{eom}
&&\!\!\!\!\!\!\!\!\!\!\!\!\!\!\!
e^{-1}\partial_{\mu}(ee_A^{\rho}S_{\rho}{}^{\mu\nu})(f_{T}+1)
 +
e_A^{\rho}S_{\rho}{}^{\mu\nu}\partial_{\mu}({T})f_{TT}
\nonumber -(f_{T}+1)e_{A}^{\lambda}T^{\rho}{}_{\mu\lambda}S_{\rho}{}^{\nu\mu}+\frac{1}{4} e_ { A
} ^ {
\nu
}[T+f({T})] \nonumber \\
&&= 4\pi Ge_{A}^{\rho}
\left[{\mathcal{T}^{(m)}}_{\rho}{}^{\nu}+{\mathcal{T}^{(r)}}_{\rho}{}^{\nu}\right], 
\end{eqnarray} 
(where $f_{T}=\partial F/\partial T$ and $f_{TT}=\partial^{2} f/\partial T^{2}$
with ${\mathcal{T}^{(m)}}_{\rho}{}^{\nu}$ and  ${\mathcal{T}^{(r)}}_{\rho}{}^{\nu}$
representing the energy-momentum tensors corresponding to matter and radiation, respectively.)\\
Next, we focus on a homogeneous and isotropic geometry, considering the standard choice of vierbeins in the analysis
\begin{equation}
  {e_{\mu}}^{A}=\textmd{diag}\left(1,a(t),a(t),a(t)\right).
\label{tetrad}
\end{equation}
This is a flat Friedmann-Robertson-Walker (FRW) background metric. We adopt the diagonal tetrad \( e^A_\mu = \mathrm{diag}(1, a(t), a(t), a(t)) \), which corresponds to a spatially flat FLRW metric. Although this tetrad is commonly used in cosmological applications of \( f(T) \) gravity, it is important to emphasize that we choose it because it satisfies the antisymmetric part of the field equations when the inertial spin connection is set to zero. This ensures the consistency of the pure tetrad formulation used throughout this work. We note that other tetrads can also reproduce the same metric, but may require a non-zero spin connection to maintain local Lorentz invariance.
\\
When the vierbein (\ref{tetrad}) is inserted into the field equations (\ref{eom}) and the Friedmann equations are obtained as:
\begin{eqnarray}\label{background1}
&&H^2= \frac{8\pi G}{3}(\rho_ {\rm m}+\rho_{\rm r})
+\frac{Tf_T}{3}-\frac{f}{6},\\\label{background2}
&&\dot{H}=-\frac{4\pi G(\rho_{\rm m}+p_{\rm m}+\rho_{\rm r}+p_{\rm r})}{1+2Tf_{TT}+f_{T}},
\end{eqnarray}
where derivatives with respect to cosmic time $t$ are indicated by dots, and $H\equiv\dot{a}/a$ is the Hubble parameter. In the aforementioned relationships, we have utilized
\begin{equation}
T=-6 H^2, \label{eq:Torsion_sc}
\end{equation}
it is directly generated for a FRW Universe by (\ref{2})\\

The structure of the first Friedmann equation (\ref{background1}) leads us to conclude that in $f(T)$ cosmology we acquire an effective dark energy sector of gravitational origin.  In particular, the density and pressure of effective dark energy can be described as
\begin{eqnarray}
&&\rho_{DE}\equiv\frac{3}{8\pi
G}\left[\frac{Tf_T}{3}-\frac{f}{6}\right], \label{rhoDDE}\\
\label{pDE}
&&P_{DE}\equiv\frac{1}{16\pi G}\left[\frac{f
+2T^2f_{TT}-f_{T} T}{1+2Tf_{TT}+f_{T}}\right],
\end{eqnarray}

consequently, its effective equation-of-state parameter is written as
\begin{eqnarray}
\label{wfT}
 w_{DE}
=-\frac{f/T-f_{T}+2Tf_{TT}}{\left[1+f_{T}+2Tf_{TT}\right]\left[f/T-2f_{T}
\right] }.
\end{eqnarray}
In the following, any quantity that has a sub-index of zero connected to it indicates its current value. We aim to challenge the model with observational data in this effort. Therefore, first we define:

\begin{eqnarray}
\label{THdef3}
E^{2}(z)\equiv\frac{H^2(z)} {H^2_{0}}=\frac{T(z)}{T_{0}},
\end{eqnarray}
with $T_0\equiv-6H_{0}^{2}$.

Therefore, using additionally that
 $\rho_{\rm r}=\rho_{\rm r0}(1+z)^{4}$,$\rho_{\rm m}=\rho_{\rm m0}(1+z)^{3}$, we
re-write the first Friedmann equation (\ref{background1}) as \cite{ref54}
\begin{eqnarray}
\label{Mod1Ez}
\frac{H^2(z,{\bf r})}{H^2_{0}}=(1+z)^3\Omega_{\rm m0}+(1+z)^4\Omega_{\rm r0}+\Omega_{\rm F0} y(z,{\bf r}),
\end{eqnarray}
where
\begin{equation}
\label{LL}
\Omega_{\rm F0}=1-\Omega_{\rm r0}-\Omega_{\rm m0} \;.
\end{equation}
In this case, $\Omega_{\rm i0}=\frac{8\pi G \rho_{\rm i0}}{3H_0^2}$ is the current density parameter for the $i$th component .  The changes to the traditional cosmological model in the context of $f(T)$ gravity are contained in the function $y(z,{\bf r})$, normalized to one at the present time.  According to the discussion in \cite{ref54}, this function is dependent on the current values of the matter and radiation density parameters, $\Omega_{\rm m0}$ and $\Omega_{\rm r0}$, as well as the parameters $r_1, r_2, \dots$ that define the particular form of $f(T)$.

\begin{equation}
\label{distortparam}
 y(z,{\bf r})=\frac{1}{T_0\Omega_{\rm F0}}\left(f-2Tf_T\right).
\end{equation}

We note that in the effective Friedman equation (\ref{Mod1Ez}), the extra component (\ref{distortparam}) is a function of the Hubble parameter only because of (\ref{eq:Torsion_sc}).

\subsection{SCALAR PERTURBATION IN $f(T)$ GRAVITY}
We discuss how the linear scalar perturbations change while considering $f(T)$ gravity. The vierbein's most prevalent linear scalar perturbations can be expressed as\cite{ref55}

\begin{eqnarray}
e^0_0 & = & a(\tau)\cdot\left(1+\psi\right),\\
e^0_i & = & a(\tau)\cdot \partial_i \zeta ,\\
e^a_0 & = & a(\tau)\cdot \partial_a \zeta,\\
e^a_j & = &a(\tau) \cdot \left((1-\phi)\delta^a_j+\epsilon_{ajk}\partial_k s\right),
\end{eqnarray}

In the Newtonian gauge, the scalar components of metric perturbations are characterized by two gravitational potentials, denoted as $\phi$ and $\psi$. The quantity $(s)$ represents the scalar contribution from spatial rotations, whereas $\zeta$ corresponds to the scalar component arising from Lorentz boosts. The symbol $\tau$ stands for conformal time. Under the Newtonian gauge choice, we set $\zeta = 0$ and $s = 0$, which simplifies the perturbations to:
\begin{align*}
e^0_0 &= (1 + \psi)\, a(\tau), \\
e^0_i &= 0, \\
e^a_0 &= 0, \\
e^a_j &= (1 - \phi)\, \delta^a_j\, a(\tau).
\end{align*}

A comprehensive discussion of linear perturbations—including scalar, vector, and tensor types—can be found in \cite{ref56}. This study specifically focuses on both scalar and tensor perturbation modes. By analyzing the time-space components of the perturbed Einstein equations, one can derive the Poisson equation as it first emerges in the context of General Relativity as:

\begin{equation}
\label{Poisson}
- k^2 \phi = 4 \pi G a^2 \sum \rho_i \Delta_i,
\end{equation}

whereas the anisotropic space-space component produces

\begin{equation}
\label{ij_term}
- k^2(\psi-\phi)  = 12 \pi G a^2 \sum  (1+w_i) \rho_i\sigma_i,
\end{equation}

where $\sigma_i$ is the anisotropic shear stress, $\delta_i$
is the density contrast, $\theta_i$ is the divergence of the peculiar velocity, and $\Delta_i= \delta_i + 3 \mathcal{H} (1+w_i) \theta_i/k^2 $ is the rest-frame density perturbation of matter species $i$. The function $\mathcal{H}$ is the Hubble function in the conformal time, and $\mathcal{H}= aH$ is the relationship between it and cosmic time. Here, prime refers to the derivative in relation to conformal time. The lensing equation is expressed as follows in GR

\begin{equation}
\label{lensing}
- k^2(\psi+\phi)  = 8 \pi G a^2 \sum \rho_i \Delta_i.
\end{equation}

For the $f(T)$ gravity, we now get the same equation. We get the Poisson equation by combining the symmetric and mixed components of the perturbation equations given in \textcolor{red}{\cite{ref6}}.

\begin{eqnarray}
\label{poisson}
-k^2 \phi &=& 4 \pi \mu_T G a^2 \sum \rho_i \Delta_i,
\end{eqnarray}
whereas the off-diagonal components and the spatial portion provide

\begin{eqnarray}
\label{lensing}
-k^2\left(\psi-R \phi\right)&=&12\pi \mu_T G a^2  \sum \rho_i (1+w_i) \sigma_i + 12 \pi \mu_T G \Xi a^2 \mathcal{H} \sum \rho_i (1+w_i) \theta_i.
\end{eqnarray}

The following quantities are defined in the above equations

\begin{equation}
\mu_T = \frac{1}{f_T},
\end{equation}

\begin{equation}
\Xi = 12 (\mathcal{H}'-\mathcal{H}^2) \mu_T f_{TT},
\end{equation}

\begin{equation}
R = 1+\frac{3 \Xi}{k^2 a^2}\left[\mathcal{H}^2 \Xi + (\mathcal{H}'-\mathcal{H}^2) a^2 \right].
\end{equation}

We recover GR for $\mu_T=1$ and $\Xi=0$, as predicted.
In $f(T)$ gravity, the lensing equation can still be expressed as

\begin{eqnarray}
-k^2\left(\psi+\phi\right) &=& 8 \pi \mu_T G a^2 S  ,\label{lensing_ft2}
\end{eqnarray}
where
\begin{eqnarray}
S &=& \sum \rho_i \Delta_i \Big[ \frac{(1+R)}{2} + \frac{3}{2}\frac{ \Xi \mathcal{H} \sum \rho_i (1+w_i) \theta_i}{\sum \rho_i \Delta_i} \Big].
\end{eqnarray}
Here we can identify

\begin{eqnarray}
\Sigma_T = \frac{\mu_T}{2} \Big[ (1+R) + \frac{3 \Xi \mathcal{H} \sum \rho_i (1+w_i) \theta_i}{ \sum \rho_i \Delta_i} \Big],
\end{eqnarray}
to define the lensing equation in its usual form

\begin{eqnarray}
\label{lensing_equation}
-k^2\left(\psi+\phi\right) &=& 8 \pi \Sigma_T G a^2 \sum \rho_i \Delta_i \label{lensing_ft3}.
\end{eqnarray}

As is customary in the literature, assuming $i=m$ (just matter contributions, ignoring neutrinos and radiation), we obtain

\begin{eqnarray}
-\frac{k^2}{a^2}\left(\psi+\phi\right) = 8 \pi \Sigma_T G \rho_m \Delta_m \label{lensing_ft4},
\end{eqnarray}
where

\begin{eqnarray}
\label{sigma}
\Sigma_T = \frac{\mu_T}{2} \Big[ (1+R) + \frac{3 \Xi \mathcal{H}  \theta_m}{ \Delta_m} \Big].
\end{eqnarray}

The aforementioned paradigm translates the deviations from GR into two phenomenological functions, the light deflection parameter $\Sigma$ and the effective gravitational coupling $\mu$, which enter the lensing and Poisson equations, respectively. The deviations from the gravitational interaction in the clustering of matter with respect to $\Lambda$CDM are encoded by the function $\mu$, while the deviation of the gravitational potential of the lens is measured by $\Sigma$. The recovery of the $\Lambda$CDM model occurs when $\mu = \Sigma = 1$.The most varied MG scenario proposals have been effectively investigated using this methodology (for a brief overview, see \cite{ref57,ref58,ref59,ref60,ref61,ref62,ref63,ref64,ref65,ref66,ref67,ref68,ref69,ref70} ).\\
The tensor perturbations modes are also considered in order to complete the set of linear perturbations. [63] 

\begin{equation}
\label{tensor}
h_{ij}^{\prime\prime}+2{\cal H}\left(1 + \frac{\Xi}{2a^2}\right)h_{ij}^{\prime}+k^2 h_{ij}=0.
\end{equation}

	Although the tensor perturbation equation~\eqref{tensor} is not explicitly used in our subsequent analysis, 
	we include it here to present the complete set of linear perturbations in $f(T)$ gravity. It is important to emphasize that the requirement $f_T < 0$ ensures the positivity of the kinetic term for tensor modes and thus avoids ghost instabilities in the gravitational sector. 
	In the present work, we implicitly assume this condition is satisfied throughout.
One crucial feature is that GWs travel at the same pace as GR. A few direct effects of GWs on $f(T)$ gravity have already been studied in \cite{ref47,ref71,ref72,ref73,ref74,ref75,ref76,ref77,ref78,ref79,ref79a,ref79b,
	ref79c,ref79d,ref79e,ref79f}.\\
In this study, we examine the power-law model provided by

\begin{equation}
f(T)= \alpha(-T)^b,
\label{36}
\end{equation}

	It is worth noting, however, that when $b < 0$, the expression $(-T)^b$ becomes singular at $T = 0$, which corresponds to Minkowski spacetime. This implies that the theory does not admit flat spacetime as a vacuum solution, raising theoretical concerns. As $T=0$ leads to divergence in the action, the Minkowski background is excluded from the solution space for negative $b$. Nonetheless, in the present work, our analysis is restricted to cosmologically relevant backgrounds, where $T$ is strictly negative due to the expanding FLRW universe. Hence, while this singular behavior at $T=0$ poses a theoretical limitation, it does not affect the dynamical evolution or the observational constraints discussed in this study.

This parametric form is arguably the most practical model and has been studied in the literature. where two model parameters, $\alpha$ and $b$, are present\cite{ref80}. When $b=0$ in \eqref{36} it reduces to the standard $\Lambda$CDM model, which provides a baseline for comparison with our modified $f(T)$ formulation.
 This parametric form is arguably the most practical model and has been studied the in the literature. Our main objective is to introduce new perspectives and methodologies for exploring the 
$f(T)$ gravity framework, rather than conducting an in-depth analysis of specific models or parameterizations. By substituting this $f(T)$ expression into the Friedmann equation at the present epoch, we obtain a theoretical constraint on the parameter $\alpha$:
\begin{equation}
\alpha=(6H_0^2)^{1-b}\frac{\Omega_{F0}}{2b-1}.
\end{equation}

This means that only one free parameter, $b$, is needed to fully specify the entire theory. We use the same methods as in\cite{ref54} to obtain the $H(z)$ function, which is the expansion rate of the universe.\\


\section{DATA AND METHODOLOGY}

The following datasets are used to derive constraints on the model baseline.
\subsection{COSMIC CHRONOMETER}
In order to measure $dz/dt$, which yields $H(z)$, the cosmic chronometers (CC) method uses the relative ages of the most massive and slowly evolving galaxies. The most recent implementation, as well as the analysis of potential sources of uncertainty, has been thoroughly detailed in \cite{ref81}. The Hubble parameter measurements compilation provided in \cite{ref81,ref82} is examined in this work. It includes the most recent updated list of $H(z)$ measurements \cite{ref81,ref82,ref83,ref84,ref85} derived using the cosmic chronometers approach, which consists of $33$ measurements spanning the redshift range $0 < z < 2$. Approximately $10$ Gyr of cosmic time are covered by this sample.
\subsection{BIG BANG NUCLEOSYNTHESIS}
In this study, we adopt the latest assumptions for Big Bang Nucleosynthesis (BBN), which include the deuterium abundance ratio, $y_{DP} = 10^5 n_D/n_H$, as reported in \cite{ref86}, along with the primordial helium fraction, $Y_P$, from \cite{ref87}. The BBN likelihood is influenced by the constraints on the effective number of neutrino species, $N_{\rm eff}$, and the physical baryon density, $\omega_b \equiv \Omega_bh^2$. Here, we assume $N_{\rm eff} = 3.046$.
\subsection{DESI BAO}
The recently published BAO measurements from the DESI collaboration incorporate data from six different tracers: quasars (QSO), Lyman-$\alpha$ forest quasars (Lya QSO), the bright galaxy survey (BGS), luminous red galaxies (LRG), emission line galaxies (ELG), and a combined LRG+ELG sample. These tracers provide estimates for the transverse comoving distance $D_M(z)/r_d$, the Hubble distance $D_H(z)/r_d$, and the volume-averaged distance $D_V(z)/r_d$ across a redshift range of $z \in [0.1, 4.2]$.
\begin{subequations}\label{desidist}
    \begin{align}
        &\frac{D_M(z)}{r_d}=\frac{D_L(z)}{r_d}(1+z)^{-1},&\\
        &\frac{D_H(z)}{r_d} = \frac{c}{r_d H(z)},&\\
        &\frac{D_V(z)}{r_d} = \frac{1}{r_d}\left[\frac{cz D^2_L(z)}{(1+z)^2H(z)}\right]^{1/3},&
    \end{align}
\end{subequations}
where $r_d$ is the drag epoch's comoving sound horizon, which we take to span the interval at steps of $2$ Mpc \cite{ref88}.

\subsection{TYPE Ia SUPERNOVAE AND CEPHEID}
Type Ia supernovae (SNe Ia) were frequently used as an important astrophysical instrument in the development of the standard cosmological model.  The Pantheon sample's Type Ia Supernovae distance moduli measurements restrict both the uncalibrated luminosity distance $H_0d_L(z)$ and the slope of the late-time expansion rate (which constrains $\Omega_m$) \cite{ref89}. \\

At redshift $z$, the theoretical apparent magnitude $m_B$ for an SN is as follows
\begin{eqnarray}
\label{distance_modulus}
m_B = 5 \log_{10} \left[ \frac{d_L(z)}{1 Mpc} \right] + 25 + M_B,
\end{eqnarray}
where $M_B$ is the absolute magnitude. The parameter $M_B$ is expected to remain constant with respect to redshift $z$, implying that the calibrated absolute magnitude of Type Ia Supernovae is typically treated as a true constant. However, it has been proposed that variations in Newton's gravitational constant G could lead to changes in the absolute luminosity, given that $L \sim 10^{-2M_B/5}$ and, conversely, the absolute magnitude $M_B$. This connection arises because the Chandrasekhar mass, which governs the luminosity of Type Ia Supernovae, depends on $G$, the Chandrasekhar mass $L \sim M_{\rm Chandra}$, which depends on $L \sim 10^{-2M_B/5}$.Therefore, a change in the strength of gravity implies an effective gravitational constant $G_{\rm eff}$, which would naturally alter the observed distance modulus. Considering equation (\ref{distance_modulus}) and a time-varying $G_{\rm eff}$,the theoretical distance modulus, which is defined by $\mu_{\rm{th}}(z) = m_B - M_B$ is modified accordingly \cite{ref69,ref70}.

\begin{equation}
\label{mb_ft}
 \mu_{\rm{th}}(z) =  5 \log_{10} d_L(z) + 25 + \frac{15}{4} \log_{10} \frac{G_{\rm eff}(z)}{G}.
\end{equation}

Taking the quasi-static approximation and the modified Poisson equation in $f(T)$ gravity context, we have \cite{new1,new2}

\begin{equation}
\frac{G_{\rm eff}(z)}{G} = \frac{1}{f_T} = \mu_T.
\end{equation}

Thus, it is possible that redshift dependence may carry useful information about the robustness of the determination of $H_0$ using the Pantheon sample and about possible modifications of $G_{\rm eff}(z)$ induced from the $f(T)$ gravity framework. 
To place constraints on the proposed scenarios, we modify the widely used and efficient cosmological tool Monte Python \cite{ref90}. The convergence of the MCMC chains is assessed using the Gelman-Rubin diagnostic $R - 1$ \cite{ref91}, ensuring it falls below the threshold of $0.01$. To assess the convergence of the MCMC chains, we employed the Gelman-Rubin diagnostic, which compares the variance within each chain to the variance between chains. A value close to $1$ indicates convergence. This diagnostic is widely used in cosmological parameter estimation due to its effectiveness in identifying non-converged or poorly mixed chains.
\vspace{0.9cm}

\section{RESULTS AND DISCUSSIONS}

Figure 1 illustrates the theoretical prediction for the Universe’s late-time expansion rate across a realistic range of values for the parameter $b$. 
Additionally, we provide the parametric space for the $f(T)$ Power-law model using latest observational datasets, including DESI BAO+BBN+Pantheon Plus, BAO+BBN+Pantheon Plus$\&$SH0ES, and  BAO+BBN+Pantheon Plus$\&$SH0ES+CC,at $68\%$ CL and $95\%$ CL in Fig. 2
\begin{figure*}[hbt!]
    \centering
    \includegraphics[width=10.9cm]{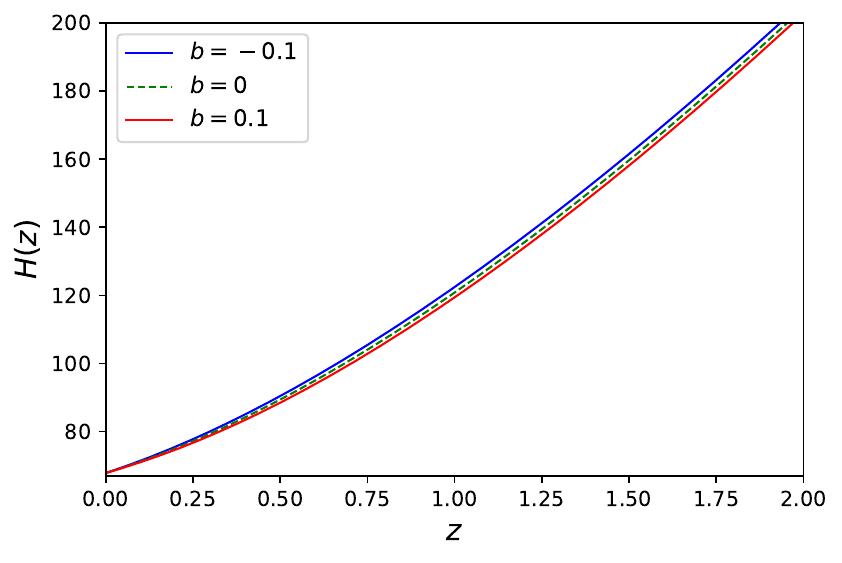}

    \caption{$H(z)$ vs $z$ $f(T)$ Model  }
   \label{H(z) vs z}
\end{figure*}
\begin{figure*}[hbt!]
    \centering
    \includegraphics[width=18cm]{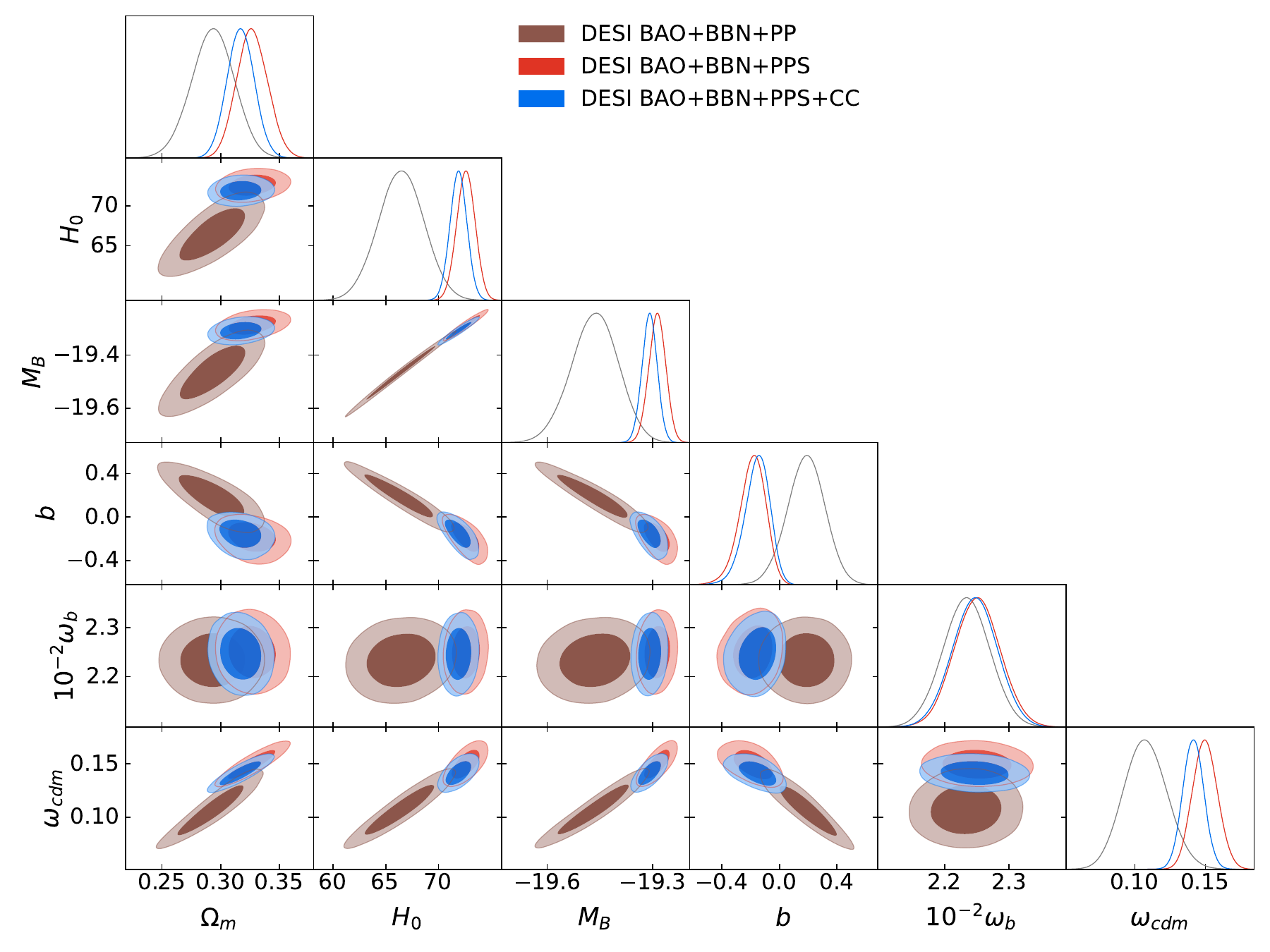}
 \caption{One-D posterior distributions and Two-D marginalized confidence regions ($68\%$ CL and $95\%$ CL) for $b$, $\Omega_{\rm m}$, $M_B$, $\omega_{cdm}$, $10^{-2}\omega_{b}$ and $H_0$ obtained from the BAO+BBN+Pantheon Plus, BAO+BBN+PantheonPlus$\&$SH0ES and BAO+BBN+Pantheon Plus$\&$SH0ES+CC for the $f(T)$ Power-law model. The parameter $H_0$ is in units of $km/s/Mpc$.  }
   \label{H(z) vs z}
\end{figure*}

\begin{table*}
\caption{Constraints at 68\% CL  on  some selected parameters of the $f(T)$ model and $\Lambda$CDM model obtained from different combination of datasets.}
\vspace{0.7cm}
\label{tableI}
 \resizebox{\textwidth}{!}{
\begin{tabular} { |c | c| c| c|l|l|  }   
 \hline

Data

 & DESI BAO+BBN+PantheonPlus &   DESI BAO+BBN+PantheonPlus\&SH0ES  & DESI BAO+BBN+PantheonPlus\&SH0ES+CC \\

 \hline
 Model      & $f(T)$ Model  &  $f(T)$ Model    & $f(T)$ Model             \\
      & \textcolor{magenta}{$\Lambda$CDM}  &  \textcolor{magenta}{$\Lambda$CDM}     & \textcolor{magenta}{$\Lambda$CDM}       \\
\hline

$b$     &        $0.18\pm0.13      $ &$-0.189^{+0.10}_{-0.079}    $ & $-0.157^{+0.096}_{-0.075}   $  \\

    &  \textcolor{magenta}{0}
& \textcolor{magenta}{0}    &  \textcolor{magenta}{0}
\\

\hline
$H_0\,[{\rm km}/{\rm s}/{\rm Mpc}]$&             $    66.5\pm2.2          $ &$   72.59\pm 0.87        $ &  $71.89\pm 0.80              $ 
\\

    &  \textcolor{magenta}{$69.24\pm 0.79$}
& \textcolor{magenta}{$71.20\pm 0.61$}    &  \textcolor{magenta}{$70.80\pm 0.60 $}
\\

\hline

$\Omega_{\rm m}$          & $  0.293\pm0.018            $ &$     0.327\pm 0.013    $ &  $   0.317\pm 0.012           $ 
\\

    &  \textcolor{magenta}{$0.312 \pm 0.012 $}
& \textcolor{magenta}{$0.319\pm 0.013$}    &  \textcolor{magenta}{$ 0.312\pm 0.011$}
\\[1ex]
\hline

$M_B$&     $  -19.465\pm0.066           $ &$   -19.287\pm 0.024       $ &  $  -19.309\pm 0.022           $ 
\\

   &  \textcolor{magenta}{$-19.382\pm 0.026 $}
& \textcolor{magenta}{$-19.320\pm 0.019$}    &  \textcolor{magenta}{$ -19.334\pm 0.018 $}
\\
\hline

$\omega_{cdm}$&     $  0.108\pm0.015            $ &$   0.1498\pm 0.0088       $ &  $  0.1414\pm 0.0074           $ 
\\

   &  \textcolor{magenta}{$0.1271^{+0.0067}_{-0.0077} $}
& \textcolor{magenta}{$0.1390\pm 0.0073$}    &  \textcolor{magenta}{$ 0.1336\pm 0.0060 $}
\\
\hline

$10^{-2}\omega_{b}$&     $  2.234\pm0.036           $ &$   2.250\pm 0.036       $ &  $  2.247\pm 0.035           $ 
\\

   &  \textcolor{magenta}{$2.233\pm 0.036 $}
& \textcolor{magenta}{$2.265\pm 0.035$}    &  \textcolor{magenta}{$ 2.260\pm 0.034 $}
\\
\hline

 


\end{tabular}
}
\label{tab:all}
\end{table*}

Table I presents the statistical results obtained from analyzing the $f(T)$ Power-law model using various observational datasets, such as DESI BAO+BBN+Pantheon Plus, BAO+BBN+Pantheon Plus$\&$SH0ES, and BAO+BBN+Pantheon Plus$\&$SH0ES+CC. Additionally, for a direct comparison between the $f(T)$ Power-law model and the standard $\Lambda$CDM model, we also include the corresponding constraints on the $\Lambda$CDM framework using the same datasets in Table I. The presence of SH0ES Cepheid host distance measurements helps to somewhat reduce the bimodal behavior observed in the parameter space. However, this bimodality still allows the parameter $b$ to align with the null hypothesis, which assumes $b = 0$. Therefore, we choose to leave the parameter $b$ free within a broad prior range, as there is no compelling reason to enforce a restrictive prior condition like $b > 0$ or $b < 0$.Since one of the primary objectives for this work is the $f(T)$ gravity model, the incorporation of SNe Ia and Cepheid host distance data from the SH0ES team is now taken into consideration along with the additional corrections on the distance moduli, i.e., \rm (\ref{mb_ft}). Since the estimate of $H_0$ from the SH0ES collaboration is derived directly from the estimate of MB, the strain on $H_0$ should be substituted with the tension on the absolute magnitude of the supernova $M_B$, as stated in Refs. [102–104]. So, in our analysis, we first consider the uncalibrated supernovae sample, i.e., the Pantheon Plus. When analyzing with BAO+BBN+Pantheon Plus, we find that ``$b > 0$ at $68\%$ CL $(b = 0.18\pm0.13 )$" for the $f(T)$ Power-law model, while we analyzing with DESI BAO+BBN+Pantheon Plus$\&$SH0ES(BAO+BBN+Pantheon Plus$\&$Sh0ES+CC) we obtained that ``$b < 0$, i.e.,  $b = -0.189^{+0.10}_{-0.079}(-0.157^{+0.096}_{-0.075})$ at $68\%$ CL". The constraints on $H_0$ from BAO+BBN+Pantheon Plus $\&$SH0ES for the $f(T)$ Power-law model are fully compatible with the estimates of $H_0$ obtained from BAO+BBN+Pantheon Plus.\\

Furthermore, the basic 1 D tension metric can be used to measure the degree of tension between two estimates $H_{0,i}$ and $H_{0,j}$ of $H_0$. It can be constructed as
\begin{equation}
T_{H_0} \equiv \frac{|H_{0,i}-H_{0,j}|}{\sqrt{\sigma^2_{H_{0,i}}+\sigma^2_{H_{0,j}}}}\,,
\label{eq:1D_estimator}
\end{equation}
\begin{figure*}[hbt!]
    \includegraphics[width=8.5cm]{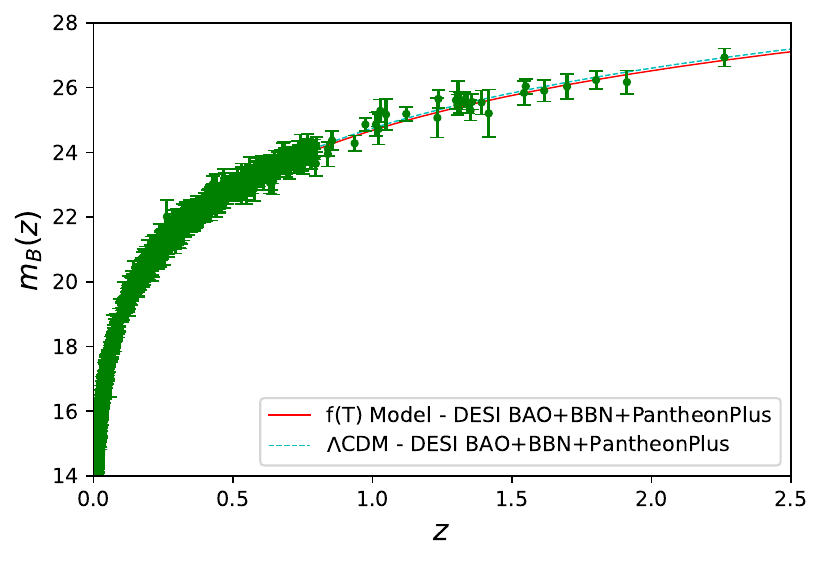}
    \includegraphics[width=8.5cm]{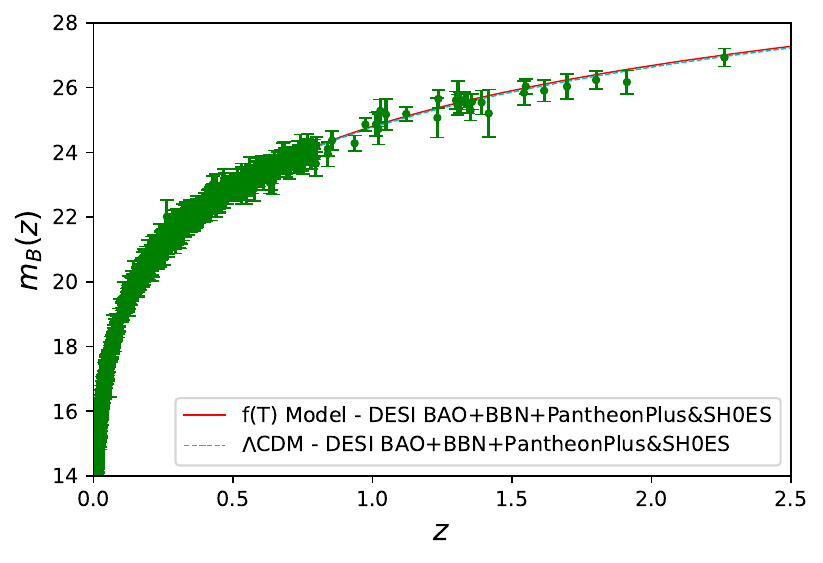}
    \includegraphics[width=8.5cm]{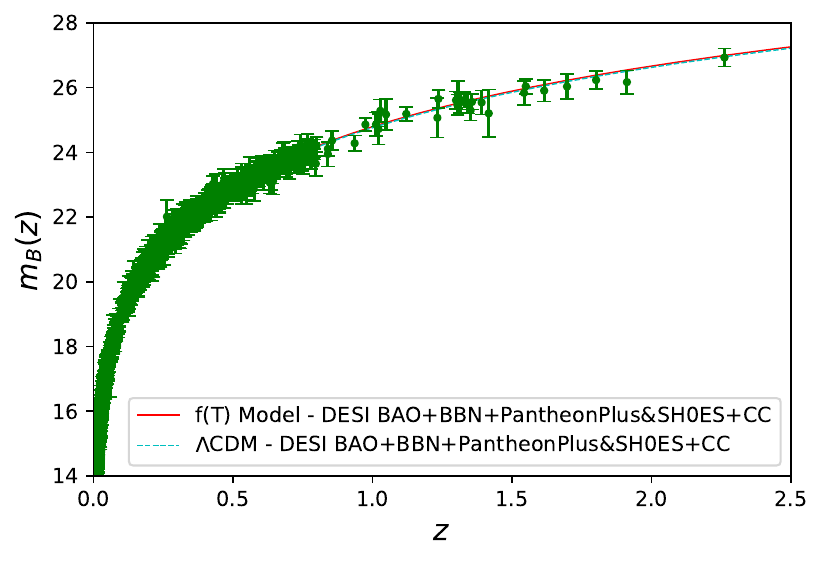}
    \includegraphics[width=8.5cm]{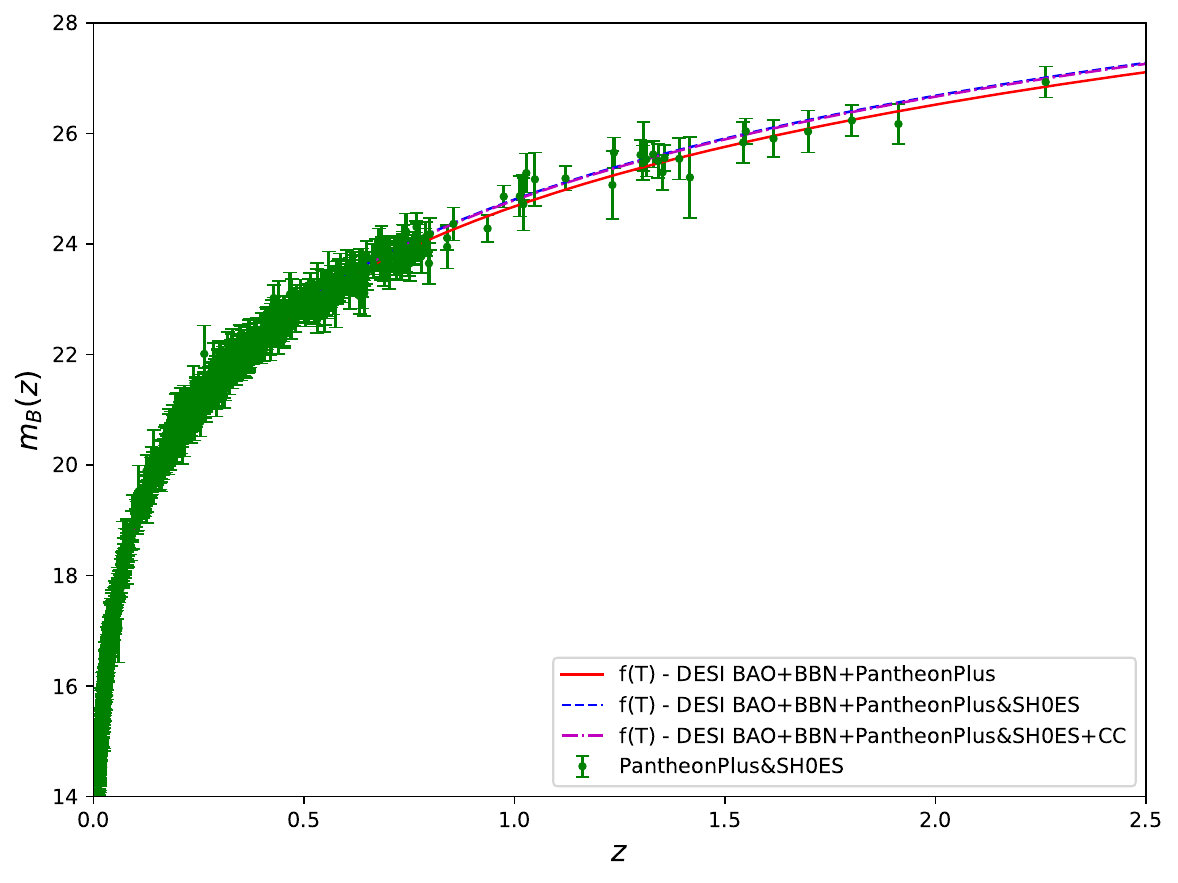}
   
    \caption{The figure illustrates the magnitude-redshift relation of the Pantheon+ dataset within the redshift range $0 < z < 2.3$. It presents the results corresponding to the best-fit parameters obtained from the combined analyses of BAO+BBN+Pantheon Plus, BAO+BBN+Pantheon Plus$\&$SH0ES, and BAO+BBN+Pantheon Plus$\&$SH0ES+CC, as listed in Table I, for the $f(T)$ Power-law model. Additionally, the prediction from the $\Lambda$CDM model using its best-fit parameters is also depicted in both panels.}
   \label{mB_ft}
\end{figure*}
The results are presented in terms of equivalent Gaussian standard deviations. Specifically, in the context of the $f(T)$ Power-law model, we observe that the $H_0$ value derived from the BAO+BBN+Pantheon Plus$\&$SH0ES datasets show a tension of $2.5\sigma$ and $0.6\sigma$ when compared to the $H_0$ values obtained from the BAO+BBN+Pantheon Plus and BAO+BBN+Pantheon Plus$\&$SH0ES+CC datasets, respectively. Furthermore, our analysis reveals a negative correlation between $H_0$ and the parameter $b$ within the $f(T)$ model across all combinations of the datasets considered (BAO+BBN+Pantheon Plus, BAO+BBN+Pantheon Plus$\&$SH0ES, and BAO+BBN+Pantheon Plus$\&$SH0ES+CC), as illustrated in Fig. 2. Consequently, an increase in $H_0$ corresponds to a decrease in $b$. Comparable findings are also obtained when considering the $M_B$ estimation approach.\\
In this study the Fig.3  illustrate the magnitude-redshift relationship for the Pantheon+ sample based on the $f(T)$ gravity model , along with the standard $\Lambda$CDM model.The plots correspond to the best-fit parameter values obtained from the combinations:
BAO+BBN+Pantheon Plus, BAO+BBN+Pantheon Plus$\&$SH0ES, and BAO+BBN+Pantheon Plus$\&$SH0ES+CC.
The results indicate that while the model predictions are nearly identical at low redshift ($z$), slight deviations may appear at higher redshift ranges. We have also analyzes comparative analysis of the theoretical predictions of the $f(T)$ gravity model against the PantheonPlus+SH0ES compilation of Type Ia Supernovae data for the observable distance modulus  $m_{B}(z)$ as a function of redshift $z$. These models incorporate progressively more observational constraints, illustrating how the inclusion of SH0ES and CC data affects the theoretical fit. Notably, all three versions of the $f(T)$ model show excellent consistency with the supernova data across the entire redshift range  $0<z<2.5$, accurately reproducing the observed cosmic acceleration trend.The subtle spread among the curves, particularly at higher redshifts,highlights the sensitivity of the model to the chosen dataset. The inclusion of SH0ES data, which provides a strong local measurement of the Hubble constant $H_{0}$,leads to a slight upward shift  in the predicted distance modulus, suggesting a marginally faster expansion rate. The further  addition of Cosmic Chronometer data refines the prediction, especially at intermediate redshifts, without deviating significantly from the observational data.This  comparative visualization demonstrates the robustness and flexibility of the $f(T)$ model under various data combinations and helps
in  distinguishing subtle differences between fits.\\
\begin{figure*}[hbt!]
    \includegraphics[width=0.58\linewidth]{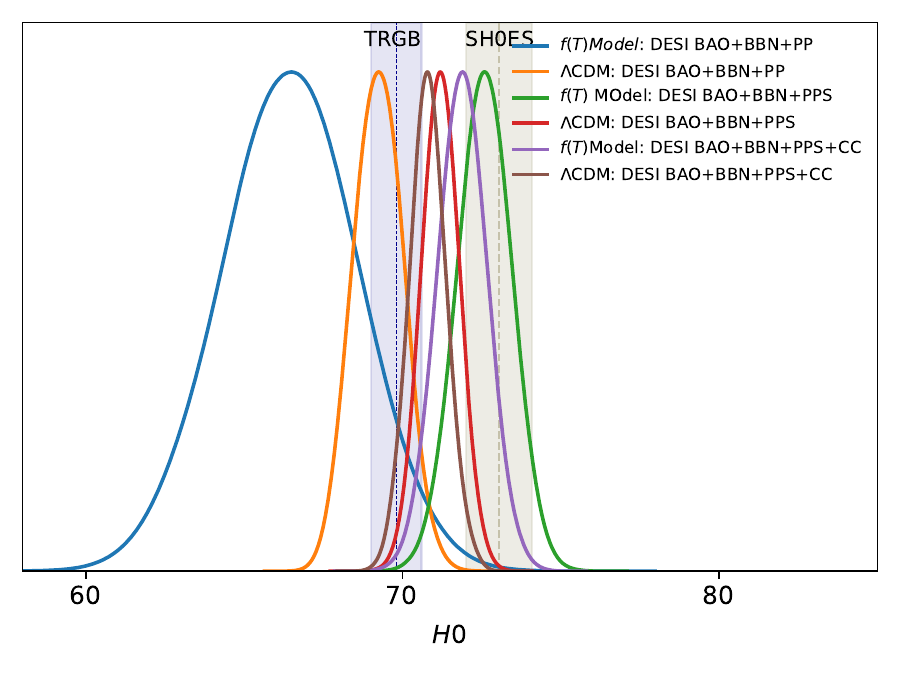}
    \caption{
    The posterior distribution functions (PDFs) for the Hubble constant $H_{0}$ are shown, with shaded areas indicating the constraint ranges and corresponding 1$\sigma$ uncertainties. These constraints are derived from SH0ES (Brout et al. 2022), Type Ia Supernovae (S.Na Ia), Big Bang Nucleosynthesis (BBN), and DESI 2024 data.}
    \label{fig:enter-label}
\end{figure*}
The Fig.4  presents posterior distribution functions (PDFs) of the Hubble constant $H_{0}$, derived from different cosmological probes. The figure shows multiple probability distributions for $H_{0}$ corresponding to different redshifts $(z)$ and observational datasets. The peak of each distribution represents the most likely value of $H_{0}$ inferred from that particular data set. The overlap and non-overlapping regions of the probability distributions highlight how different methods provide independent but sometimes conflicting measurements.  The vertical dashed lines labeled Planck, TRGB, and SH0ES represent key $H_{0}$measurements.Planck (CMB-based) typically predicts a lower $H_{0}$, while SH0ES (Cepheid-based) gives a higher, highlighting the Hubble tension.The overlap  between different distributions illustrates the degree of agreement between different methods.\\
\begin{figure*}[hbt!]
    \includegraphics[width=0.75\linewidth]{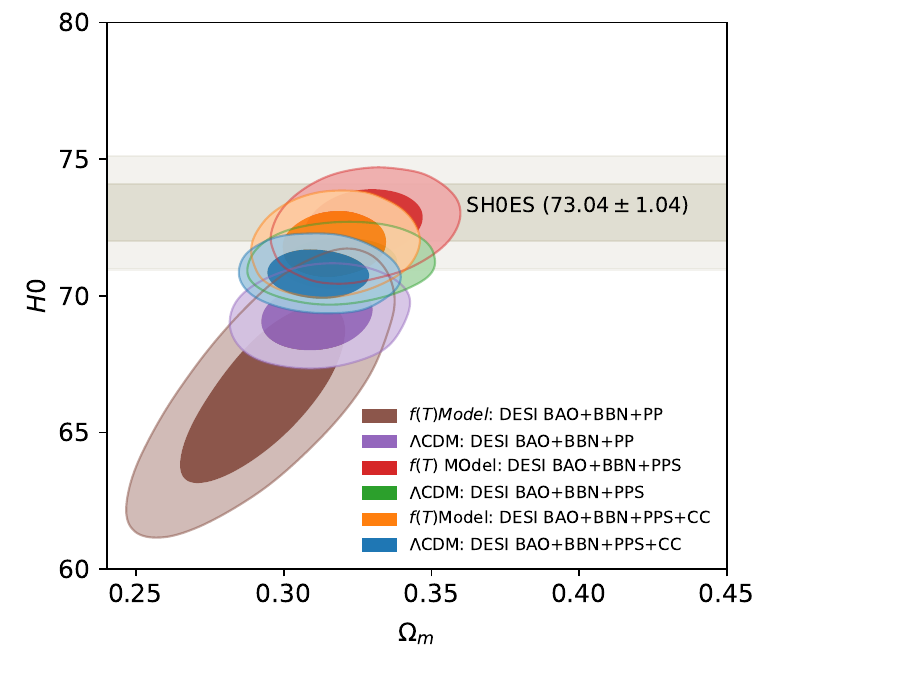}
    \caption{Two-dimensional marginalized confidence regions at 68\% and 95\% C.L of $\Omega_m$ and $H_0$ for the $w$CDM and $\Lambda$CDM 
models from all considered combination data sets are presented in Table I. The horizontal light brown band  represents SH0ES $(H_0=73.04 \pm 1.04)$ measurement}
    \label{fig:enter-label}
\end{figure*}
In this study, we examine the magnitude-redshift relationship of the Pantheon Plus sample for both the $f(T)$ gravity model and the reference $\Lambda$CDM model. This analysis uses the best-fit parameters obtained from the datasets: BAO+BBN+Pantheon Plus, BAO+BBN+Pantheon Plus$\&$SH0ES, and BAO+BBN+Pantheon Plus$\&$SH0ES+CC, as shown in Fig. 3. Our results indicate that at low redshift ($z$), the predictions of the models are nearly indistinguishable, while slight deviations may appear at higher redshift values.\\

In order to provide a comprehensive comparison of the constraints on the Hubble constant $H_{0}$ and the matter density parameter $\Omega_{m}$, we present the two-dimensional marginalized confidence contours in Fig. 5. The figure displays the 68$\%$ and 95$\%$ confidence regions for both the $f(T)$ and $\Lambda$CDM models under various combinations of observational datasets, such as DESI BAO, BBN, PPS, and Cosmic Chronometers (CC). Notably, the horizontal light brown band in the plot represents the SH0ES constraint $(H_{0}=73.041\pm1.04$, which allows for a direct visual comparison of the model predictions with local measurements. It is evident from the figure that while some combinations of the $f(t)$ model come closer to the SH0ES value of $H_{0}$ the standard $\Lambda$CDM model remains in tension, particularly for datasets excluding late-time observables. This comparison underscores the potential of extended gravity models in addressing the Hubble tension.

\section{Model Comparison}

The current subsection concludes the observational analysis by comparing the fits of the different models based on the accepted information criteria. The Akaike Information Criterion (AIC) and the Bayesian or Schwarz Information Criterion (BIC) are the two primary examples of these criteria. These are each described as
\begin{eqnarray}
&& AIC  = -2 \ln \mathcal{L}+ 2 d = \chi^2_{min} + 2 d,\label{aic}
\end{eqnarray}
and
\begin{eqnarray}
&& BIC  = -2 \ln \mathcal{L}+  d \ln N = \chi^2_{min} +  d \ln N,\label{bic}
\end{eqnarray}

Here, $\mathcal{L} = \exp\left(-\chi_{\text{min}}^2/2\right)$ represents the maximum likelihood, with $d$ indicating the number of parameters in the model and $N$ denoting the total number of data points considered in the analysis. To evaluate and compare different models, it is essential to adopt a baseline scenario, which is the standard $\Lambda$CDM cosmology. For any given model $M$, one can assess its relative performance by calculating the difference $\Delta X = X_M - X_{\Lambda\text{CDM}}$, where $X$ corresponds to either the AIC or BIC criterion. Based on the value of $\Delta X$, the model's level of support can be interpreted as follows:  
(i) Strong support if $\Delta X \leq 2$ (indicating a good fit),  
(ii) Moderate support if $4 \leq \Delta X \leq 7$, and  
(iii) Essentially no support if $\Delta X \geq 10$.  
Table II presents the $\Delta X$ values for the $f(T)$ model across three combinations of data sets: DESI BAO+BBN+PP, DESI BAO+BBN+PPS, and DESI BAO+BBN+PPS+CC. The results suggest that the $f(T)$ Power-law model slightly outperforms the $\Lambda$CDM model and demonstrates excellent consistency with observational data across all tested data sets.

\vspace{1cm}

\begin{table*}[hbt!]

\label{tableI}
 \resizebox{\textwidth}{!}{
\begin{tabular} { |c | c| c| c|l|l|  }   
 \hline

Data

 & DESI BAO+BBN+Pantheon Plus &   DESI BAO+BBN+Pantheon Plus\&SH0ES  & DESI BAO+BBN+Pantheon Plus\&SH0ES+CC \\

 \hline
 Model      & $f(T)$ Model  &  $f(T)$ Model    & $f(T)$ Model             \\
      & \textcolor{magenta}{$\Lambda$CDM}  &  \textcolor{magenta}{$\Lambda$CDM}     & \textcolor{magenta}{$\Lambda$CDM}       \\
\hline

$AIC$     &        $1437.54     $ &$1325.24    $ & $1343.94   $  \\

    &  \textcolor{magenta}{$1438.12$}
& \textcolor{magenta}{$1327.9$}    &  \textcolor{magenta}{$1345.22$}
\\

\hline
$\Delta AIC$&             $    0.55          $ &$   2.66      $ &  $1.25             $ 
\\
 
    &  \textcolor{magenta}{$0$}
& \textcolor{magenta}{$0$}    &  \textcolor{magenta}{$0$}
\\

\hline

$BIC$          & $  1475.66        $ &$   1363.68      $ &  $   1382.50         $ 
\\
 
    &  \textcolor{magenta}{$1470.76 $}
& \textcolor{magenta}{$1360.84$}    &  \textcolor{magenta}{$ 1378.27$}
\\
\hline

$\Delta BIC$&     $     4.89      $ &$  2.83      $ &  $ 4.22  $ 
\\

   &  \textcolor{magenta}{$0$}
& \textcolor{magenta}{$0$}    &  \textcolor{magenta}{$ 0 $}
\\
\hline

\end{tabular}
}

\caption{
A summary of the $AIC$ and $BIC$ values, along with their deviations from the standard $\Lambda$CDM cosmological model, is provided for various $f(T)$ gravity models. The analysis includes different data combinations: DESI BAO+BBN+PP, DESI BAO+BBN+PPS, and DESI BAO+BBN+PPS+CC.}
\vspace{0.5cm}
\label{tab:all}
\end{table*}

\section{conclusion}

In this study, we explored the framework of $f(T)$ gravity as an extension of teleparallel gravity, focusing on its cosmological implications and deviations from General Relativity (GR). By modifying the torsion scalar in the gravitational action, $f(T)$ gravity provides an avenue to explain late-time cosmic acceleration without the explicit need for a cosmological constant. We examined its impact on background evolution, linear perturbations, and the implications for structure formation and gravitational lensing. From the background evolution perspective, we used the modified Friedmann equations and demonstrated how the $f(T)$ modification leads to an effective dark energy component. We showed that the additional function $F(T)$ contributes dynamically to cosmic expansion and alters the effective equation-of-state parameter, leading to deviations from the standard $\Lambda$CDM model. Recent cosmological data from the DESI collaboration has provided new perspectives for understanding the universe. In this study, we utilize the DESI BAO, BBN, and Pantheon Plus $\&$ SHOES datasets to investigate the cosmological evolution of dynamical dark energy using a Gaussian process. We perform statistical analysis on each individual model and determine the best-fit values using the various datasets. The constraints on the functional form of $f(T)$ were also established by linking them to observational cosmological parameters such as DESI BAO, BBN, and Pantheon Plus $\&$ SHOES datasets.\\

To test the viability of the $f(T)$ framework, we adopted a power-law model parameterized by $f(T) = \alpha(-T)^b$. This form allowed us to express all modifications in terms of a single free parameter $b$, making it suitable for observational testing. Theoretical constraints on $\alpha$ were established using the present-day Friedmann equation. Using numerical methods and modifying the CLASS code, we implemented this model to compare with observational datasets such as cosmic chronometers, supernova data, and large-scale structure surveys. \\

Using the available observational data. We determined
the 2-Dimensional likelihood contours with errors of 1-$\sigma$ and 2-$\sigma$ with confidence values of 68$\%$ and 95$\%$for DESI BAO+BBN+Panthon Plus, or DESI BAO+BBN+Panthon Plus $\&$ SHOES data and Panthon Plus $\&$ SHOES data, respectively.We compared the fitting procedure of the $f(T)$ models with the $\Lambda$CDM model using the AIC and BIC information criteria. According to the AIC, the $f(T)$ model performed well and slightly outperformed the $\Lambda$CDM model various both data sets. However, based on the BIC, the $\Lambda$CDM scenario was marginally superior,though all $f(T)$ models remained quite efficient.In conclusion, by utilizing the newly released DESI data along with BAO, BBN and Pantheon $\&$ SHOES data from other probes, we performed a fitting of the most widely used
and viable $f(T)$ gravity models. As observed, $f(T)$ gravity is in strong agreement with the data.Overall, our analysis highlights that $f(T)$ gravity provides a compelling alternative to dark energy models within the framework of modified gravity. While it successfully describes late-time cosmic acceleration, further constraints from observational data, such as Cosmic Microwave Background anisotropies and baryon acoustic oscillations, are necessary to determine the parameter space where $f(T)$ gravity remains a viable alternative to $\Lambda$CDM. Future studies should focus on refining the theoretical predictions of $f(T)$ models and testing them with upcoming high-precision cosmological surveys to establish their consistency with observational data and fundamental physical principles.\\

Through this work, we aim to provide a comprehensive analysis of gravity and its implications for cosmology, with particular emphasis on background evolution, perturbations, and observational tests. Our results contribute to the growing body of literature investigating alternatives to general relativity and their potential to explain the accelerating universe without the need for an explicit cosmological constant.


\section*{Declaration of competing interest}
The authors declare that they have no known competing financial
interests or personal relationships that could have appeared to influence
the work reported in this paper.

\section*{Data availability}
No data was used for the research described in the article.

\section*{Acknowledgment}
The authors, A. Dixit and A. Pradhan, appreciate the facilities provided by Inter-University Centre for Astronomy \& Astrophysics (IUCAA), Pune, India during their visiting associateship program. The authors sincerely thank the reviewer and editor for their comments and suggestions, which improved the manuscript in its current form.

\end{document}